# EXPLAINABLE AI FOR INTERPRETABLE CREDIT SCORING

Lara Marie Demajo, Vince Vella and Alexiei Dingli

Department of Artificial Intelligence, University of Malta, Msida, Malta

## ABSTRACT

*With the ever-growing achievements in Artificial Intelligence (AI) and the recent boosted enthusiasm in Financial Technology (FinTech), applications such as credit scoring have gained substantial academic interest. Credit scoring helps financial experts make better decisions regarding whether or not to accept a loan application, such that loans with a high probability of default are not accepted. Apart from the noisy and highly imbalanced data challenges faced by such credit scoring models, recent regulations such as the `right to explanation' introduced by the General Data Protection Regulation (GDPR) and the Equal Credit Opportunity Act (ECOA) have added the need for model interpretability to ensure that algorithmic decisions are understandable and coherent. An interesting concept that has been recently introduced is eXplainable AI (XAI), which focuses on making black-box models more interpretable. In this work, we present a credit scoring model that is both accurate and interpretable. For classification, state-of-the-art performance on the Home Equity Line of Credit (HELOC) and Lending Club (LC) Datasets is achieved using the Extreme Gradient Boosting (XGBoost) model. The model is then further enhanced with a 360-degree explanation framework, which provides different explanations (i.e. global, local feature-based and local instance-based) that are required by different people in different situations. Evaluation through the use of functionally-grounded, application-grounded and human-grounded analysis show that the explanations provided are simple, consistent as well as satisfy the six predetermined hypotheses testing for correctness, effectiveness, easy understanding, detail sufficiency and trustworthiness.*

## 1. INTRODUCTION

### 1.1. Problem Definition

Credit scoring models are decision models that help lenders decide whether or not to accept a loan application based on the model's expectation of the applicant being capable or not of repaying the financial obligations [1]. Such models are beneficial since they reduce the time needed for the loan approval process, allow loan officers to concentrate on only a percentage of the applications, lead to cost savings, reduce human subjectivity and decrease default risk [2]. There has been a lot of research on this problem, with various Machine Learning (ML) and Artificial Intelligence (AI) techniques proposed. Such techniques might be exceptional in predictive power but are also known as black-box methods since they provide no explanations behind their decisions, making humans unable to interpret them [3]. Therefore, it is highly unlikely that any financial expert is ready to trust the predictions of a model without any sort of justification [4]. Model explainability has recently regained attention with the emerging area of

eXplainable AI (XAI), a concept which focuses on opening black-box models in order to improve the understanding of the logic behind the predictions [5, 6]. With regards to credit scoring, lenders will need to understand the model's predictions to ensure that decisions are made for the correct reasons. Furthermore, in adherence to existing regulations such as the GDPR 'right to explanation' and the ECOA, applicants have the right to know why they have been denied the loan. Therefore, credit scoring models must be both exceptional classifiers and interpretable, to be adopted by financial institutions [7, 8]. Formally, in this work we refer to model interpretability as the model's ability to explain or to present in understandable terms to a human [9]. The terms *explainability*, *interpretability*, *understandability* and *comprehensability* are used interchangeably in this work.

There are a number of challenges posed when working with XAI, including questions like "who are the explanations for (experts or users)?", "what is the best form of representation for the explanations?" and "how can we evaluate the results?" [10]. The literature still lacks precise answers to these questions since different people require different types of explanations. This lead to ambiguity in regulations and solutions needed [11]. The literature includes very few entrances of such interpretable credit scoring models, most of which provide only a single dimension of explainability. Therefore, in this work, we shall be addressing this gap by proposing a credit scoring model with state-of-the-art classification performance on two popular credit datasets (HELOC and Lending Club Datasets) and enhanced by a 360-degree explanation framework for model interpretability by bringing together different types of explanations.

## 1.2. Aims and Objectives

Our goal of an interpretable credit scoring model can be decomposed into the following two main objectives:

1. Model interpretability of the implemented credit scoring model by providing human-understandable explanations through different XAI techniques (Section 3.3)
2. A comprehensive approach for evaluation of model interpretability through both human subjective analysis and non-subjective scientific metrics (Section 4)

Details about how these objectives have been met are found in the rest of this paper, which is organized as follows. Chapter 2 includes a review of existing methods in the XAI domain. A detailed workflow of the system is discussed in Chapter 3. Chapter 4 includes all the experiments carried out to evaluate the interpretability performance, whilst any limitations, improvements and conclusions are finally discussed in Chapter 5.

## 2. RELATED WORK

Back in 1981, [12] state that the ability to explain decisions is the most highly desirable feature of a decision-assisting system. Recently, XAI has gained high popularity. It aims to improve the model understandability and increase humans' trust. There have been various efforts in making AI/ML models more explainable in many applications, with the most popular domain being image classification [13, 14, 15].

The authors in [16] state that dimensionality reduction like feature selection and Principle Component Analysis (PCA) can be an efficient approach to model interpretation since the outcome can be intuitively explained in terms of the extracted features. Štrumbelj and Kononenko in [17] propose a sensitivity analysis based model, which analyses how much each feature contributes to the model's predictions by finding the difference between the prediction and expected prediction when the feature is ignored. Such explanations are given in the form of feature contributions.

Trinkle and Baldwin in [18] investigate whether Artificial Neural Networks (ANNs) can provide explanations for their decisions by interpreting the connection weights of the network. They conclude that performance was restricted due to the use of just one hidden layer and state that such interpretation techniques are not robust enough to handle more hidden layers. Baesens et al. in [19] contributed to making ANNs more interpretable by making use of NN rule extraction techniques to investigate whether meaningful rule sets can be generated. They implemented three NN rule extraction techniques being Neurorule, TREPAN and Nefclass, and the extracted rules were then presented in a decision tree structure since graphical representations are more interpretable by humans.

The authors in [20] propose Layer-wise Relevance Propagation (LRP), a post-hoc interpretability model for interpreting the individual predictions of a Deep Neural Network (DNN) rather than the model itself. It propagates back through the layers of the network until reaching the input and pinpoints the regions in the input image that contributed the most to the prediction. In [21], Yang et al. propose Global Interpretation via Recursive Partitioning (GIRP), a compact binary tree that interprets ML models globally by representing the most important decision rules implicitly contained in the model using a contribution matrix of input variables. Ribeiro et al. in [22] propose Local Interpretable Model-agnostic Explanations (LIME), a novel technique that explains any classifier's predictions by approximating them locally with a secondary interpretable model. While these are local explanations, the global view of the model can be presented by selecting a set of representative and non-redundant explanations. In [23], Ribeiro et al. introduce another novel model-agnostic system to explain the behaviour of complex models. They propose Anchors, which are intuitive high precision IF-THEN rules that highlight the part of the input, which is used by the classifier to make the prediction. It is shown that Anchors yield better coverage and require less effort to understand than LIME. The authors in [24] propose SHapley Additive eXplanations (SHAP), a unified framework for interpreting predictions. SHAP are Shapley values representing the feature importance measure for a particular prediction and are computed by combining insights from 6 local feature attribution methods. Results show that SHAP are consistent with human intuition.

In 2018, Fair Isaac Corporation (FICO) issued the Explainable Machine Learning Challenge in aim of generating new research in the domain of algorithmic explainability. They challenged participants to create ML models that are both accurate and explainable, in aim of solving the credit scoring problem using the HELOC financial dataset. The final models were qualitatively evaluated by data scientists at FICO. The winners, Dash et al. [25], propose Boolean Rules via Column Generation (BRCG), a novel global interpretable model for classification where Boolean rules in disjunctive normal form (DNF) or conjunctive normal form (CNF) are learned. Column generation is used to efficiently search through the number of candidate clauses without heuristic rule mining. BRCG dominates the accuracy-simplicity trade-off in half of the datasets tested, but even though it achieves good classification performance and explainability, methods like the RIPPER decision tree still obtain a better classification accuracy in many of the datasets, including HELOC. The authors state that limitations include performance variability as well as the reduced solution quality when implemented on large datasets. In [26], Gomez et al. propose another solution to this challenge. They make use of a Support Vector Machine (SVM) model with a linear kernel for classification, where features were first discretized into ten bins to get rid of outliers and ensure scalability and manageable time complexity. For explainability, they implement an updated version of Anchors [23], which finds key features by systematically perturbing the columns and holding others fixed. They combine instance-level explanations and global-level model interpretations to create an interactive application visualising the logic behind the model's decisions, identifying the most contributing features in a decision. Using a greedy approach, they also suggest the minimal set of changes required to switch the model's output. Chen et al. are also mentioned for their great work in [27], who propose a globally interpretable model known as the 'two-layer additive risk model', achieving accuracy similar to other neural networks. The model is decomposable into subscales, where smaller models are created from subgroups of features and eventually combined to produce the final default probability. The decomposable nature of the model allows it to produce meaningful components that identify the list of factors that contribute most to the model's predictions, providing rule-based summary explanations for global interpretability and local case-based explanations for local interpretation.

All these XAI techniques present their ability in making ML/AI models more interpretable. The last three mentioned techniques are credit scoring models that provide good classification performance as well as local and/or global explainability. In fact, they present an interesting evolution of explainability within credit scoring, motivating the main goal of this project. The winner of FICO's Explainable Machine Learning Challenge in 2018, BRCG [25], is considered as a state-of-the-art of XAI in credit scoring and is therefore selected as the benchmark paper for this work.

## 3. METHODOLOGY

For this work, an interpretable credit scoring model is proposed. As depicted in Figure 1 the data is first preprocessed and then a classification function is adopted to classify the data instances. Subsequently, the classifier is extended by three XAI methods to provide a 360-degree explanation framework. Therefore, the pipeline of the system comprises of three main sequential phases:

1. **Data handling & preprocessing**: transforming and preparing data for classification
2. **Classification**: classifying data into predetermined labels
3. **eXplainable AI**: appending interpretable explanations to the classification predictions

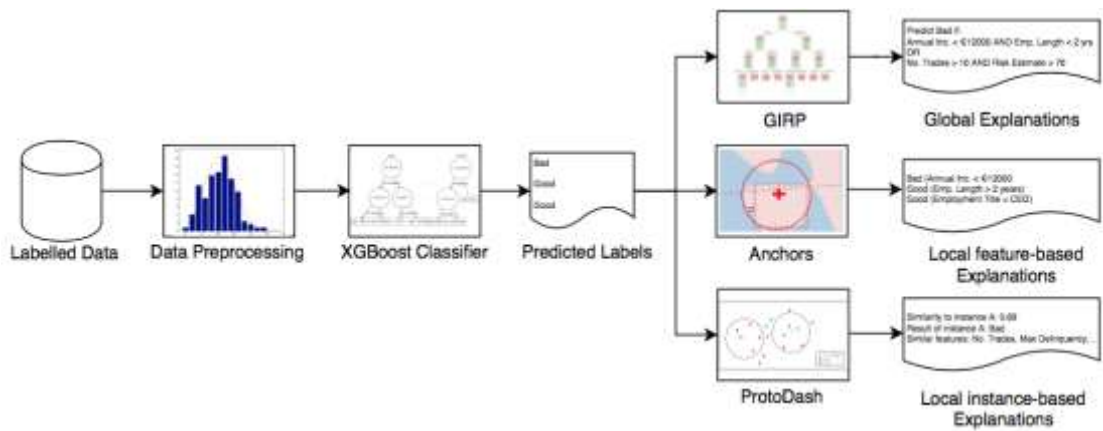

Figure 1. Pipeline of proposed Interpretable Credit Scoring model.

### 3.1. Data Handling and Preprocessing

Considering that our classifier requires supervised learning, the first concern is the preparation and handling of the data. Two datasets are used in this project:

- The Home Equity Line of Credit (HELOC) Dataset which contains around 10,000 instances with 24 different features (21 numerical and 3 categorical).
- The Lending Club (LC) Dataset which includes around 2.3 million loan applications with 145 features of different types.

The HELOC Dataset is used by the benchmark paper [25] and therefore used for fair comparison, whilst the LC Dataset is quite popular in the credit scoring literature [28, 28, 30, 31] and is used to further evaluate the implemented model. Figure 2 depicts the different stages undertaken during the preprocessing phase until the data is ready to be used by the model.

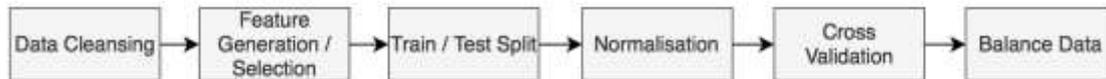

Figure 2. Pipeline of the different stages during the data preprocessing phase.

Firstly, the data is cleaned by handling any special values in the HELOC Dataset, converting the target variables into binary labels, removing outliers, imputing empty values in the numerical features with the mean and imputing empty values in categorical features with unused category values, converting categorical features using one-hot encoding, and eliminating noisy data. Next, some additional variables are computed for the Lending Club Dataset such as *Loan Amount to Annual Income* since ratios are better for deep learning classifiers, and feature selection is performed on the LC Dataset through analysis of the correlation matrix and change in classification performance. Furthermore, both datasets are split with 75:25 ratio using stratification and the training set is further split using stratified 10-fold cross validation. The data for each fold is then normalised using both the min-max normalisation and standard scaling techniques and best method is identified through evaluation. Finally, the training data of the LC Dataset is balanced using four different resampling techniques and best method is also identified through evaluation.

## 3.2. Classification

Initially, a DNN model was chosen as the classification function, however after abundant experiments and evaluation it was noted that the performance was not as satisfactory as expected. As observed by the authors in [32] and [33], DNNs often suffer from reduced performance on certain credit scoring datasets. After replicating the work in [29], who use an ANN and present a high accuracy, results show that their model was actually overfitting and classifying all instances into the same class (i.e. giving a high accuracy due to data being highly unbalanced). From the literature [32, 34, 35] it was observed that other methods like Random Forest (RF) and Extreme Gradient Boosting (XGBoost) perform strongly in credit scoring. Hence, for this work, the most commonly used classification techniques are implemented and compared in order to find the best performing algorithm in credit scoring on both the HELOC and LC Datasets. The ML models implemented include Logistic Regression, Decision Tree, Random Forest, ANN and XGBoost. A "light" version of the BRCG algorithm by Dash et al. [25], which is a more efficient variant of the method, is also implemented in this work to enable better comparison with more metrics between the classification techniques.

Results show that the XGBoost model does not require any data normalisation and that the oversampling technique gave the best performance on the Lending Club Dataset. Using the optimal parameter values found by the grid search on the HELOC Dataset portrayed classification performance improvement on both datasets. It is concluded that XGBoost yields less Type-I and Type-II errors than LBRCG on the HELOC Dataset whereas LBRCG yields less Type-I errors (5,112 loans) but much more Type-II errors (27,767 loans) than XGBoost on the LC Dataset. This signifies that the XGBoost model is better at maintaining a good balance between Type-I and Type-II errors and improves performance over the LBRCG model by an F1-Score of more than 3% on the HELOC Dataset and an F1-Score of over 7% on the LC Dataset.

### 3.3. eXplainable AI

Given their opaque nature, deep learning techniques lack interpretability and are unable to explain their decisions. Therefore, the main goal of this project is to enhance the implemented credit scoring model by augmenting it with a 360-degree explanation framework such that it can provide different types of explanations for its predictions. Many new XAI methods have been recently published in the literature, some of which have not yet been explicitly applied to the credit scoring domain.

It is important to note that different people in diverse situations require different explanations [8]. There are three different personas that require explainability in credit scoring, being (i) loan officers that are said to prefer local sample-based explanations, (ii) rejected loan applicants that are said to prefer local feature-based explanations, and (iii) regulators or data scientists that are said to prefer global model explanations. Hence, a single XAI method does not suffice to provide all the explanations required [11]. In this project, we aim to propose an explainable credit scoring model that provides 360-degree interpretability by producing explanations for each of the three different personas mentioned. Table 1 represents the three different state-of-the-art XAI methods used to yield the different explainability dimensions required. The following subsections describe the implementation details of the XAI method implemented for each of the three different explanation types.

Table 1. The XAI methods used to generate each explanation type and the format of the explanation provided by each method.

| Explanation Type | XAI Method | Explanation Form | Reference |
| --- | --- | --- | --- |
| Global | SHAP+GIRP | Decision Tree / IF-THEN rules | [24], [21] |
| Local feature-based | Anchors | DNF rule | [23] |
| Local instance-based | ProtoDash | Prototypical instances | [36] |

#### 3.3.1. Global Explanations

Global explanations are explanations that provide a global understanding of how the classification model works overall. They interpret the reasoning behind the general logic used by the model when making its predictions. Such explanations are usually preferred by regulators and managers since they are mostly concerned with the global understanding of the credit scoring model rather than the individual explanations of each instance. This is because regulators and data scientists are responsible of ensuring that the model is being correct, fair and compliant in its predictions.

As discussed previously, the benchmark BRCG method [25] is a directly interpretable supervised learning method that provides Boolean rules to globally explain its logic. Therefore, in this project, we aim to implement an XAI technique that provides similar or better global explanations to BRCG. For this part of the XAI objective, a SHAP+GIRP method is implemented. GIRP [21] is a post-hoc method that is capable of interpreting any black-box model by extracting the most important rules used by the model in its predictions. It is a very recent model that depicts state-of-the-art capabilities in model-agnostic interpretability. It is important to note that, to the best of our knowledge, GIRP has not yet been explicitly applied to the credit risk problem and no formal academic results have been presented for this domain. The explanation provided by GIRP for the global understandability of the model is given in the form of a decision tree, however the IF-THEN rules that make up the tree are also extracted and provided for variety.

GIRP makes use of a contribution matrix to generate a decision tree consisting of the most discriminative rules contained in the trained model, which is then pruned for better generalisation to form the Interpretation Tree. The contribution matrix contains the contributions of each input variable to the prediction of each instance. To generate this contribution matrix, Lundberg and Lee's SHAP [24] method is used and is implemented using the SHAP Python library. Using the SHAP package, an explainer is created over the XGBoost model to generate SHAP values for each feature for each prediction, constructing the contribution matrix for GIRP. For the implementation of GIRP, source code was adapted from a Github repository (https://github.com/west-gates/GIRP [Accessed: 10/08/2020]) containing an implementation of GIRP on text classification. The code was updated such that the methods handle tabular data rather than words from text extracts. The rules extracted from GIRP include a number of conditions in their IF statement and a default rate in their THEN section. The larger the default rate, the higher the risk.

### 3.3.2. Local Feature-based Explanations

Local explanations are explanations that provide a local understanding on how and why a specific prediction was made. Such explanations are said to be preferred by loan customers since they are mostly interested in why they have been denied the loan and what is the reasoning behind the model's prediction for their particular loan application. This type of explanation can be provided in the form of feature relevance scores or rules. It is important to note that local explanations are not provided by the benchmark BRCG model. However, in this work, we aim to go above and beyond global explainability. As discussed, Lu et al. in [8] state that different people in different scenarios require different explanations and therefore, we aim to provide further model interpretability through local explanations.

In this work, the local feature-based explanations are generated using the post-hoc Anchors method from [23]. The explanations are given in the form of rules containing conditions on the most important features for the model prediction. The original Python implementation of Anchors from [23] is used. Anchors generates an anchor rule that is iteratively increased in size until a predetermined probability threshold is reached. The outputted rule is the shortest rule with the largest coverage and closest estimated precision to the threshold, that explains the model prediction. The anchor rule contains the features and feature values that contributed to the model prediction. An anchor rule is a sufficient condition, which means that other data points that satisfy it should have, with an $x\%$ probability, the same prediction as the original data point. The probability $x$ is set to 90% in this work. It was noted that for the HELOC Dataset, the resulting anchor rule only holds for the data point it was built for (from the entire test data), even when reducing the probability threshold $x$ to 50%. Furthermore, the outputted anchor rules contain an average of 35 conditions, which might make the rule hard to read and consequently uninterpretable. Therefore, we implemented a further extension that iterates over the partial anchors in the main anchor to find the shortest partial anchor that still holds for the data point. This obtains rules that contain an average of 4 conditions. Finally, the derived rule of each data point is used as the local feature-based explanation for its prediction.

### 3.3.3. Local Instance-based Explanations

Similar to local feature-based explanations, these explanations provide a local understanding on individual predictions rather than the model as a whole. Such explanations are said to be preferred by loan officers since they are interested in validating whether the prediction given by the model for a loan application is justified. Therefore, it is said that a loan officer would gain more confidence in the model's prediction by looking at other similar loan applications with the same outcome, and hence understanding why a loan application has been denied compared to other loan applications that were previously accepted and then ended up defaulting [11]. This type of explanation is usually provided in the form of prototypes (i.e. similar data points from the dataset). Again, it is important to note that this type of local explanations is not provided by the benchmark BRCG model but in this work, we aim to provide further model interpretability by providing explanations for the three personas that require explainability in credit scoring.

In this work, local instance-based explanations are generated using the post-hoc ProtoDash method by Gurumoorthy et al. [36]. The explanations are given in the form of two prototypical data points that have similar features. The implementation of ProtoDash by AIX360 [11] is used. ProtoDash employs the fast gradient-based algorithm to find prototypes of the data point in question as well as the non-negative importance weight of each prototype. The algorithm aims to minimize the maximum mean discrepancy metric and terminates when the number of prototypes $m$ is reached. For this work, $m=6$ is used and the two prototypes with the largest weight from the outputted six are selected as the final exemplar-based explanation.

## 4. EVALUATION & RESULTS

The aim of this study is to enhance the credit scoring system with interpretability such that its predictions are also justified by reasons. However, these reasons need to make sense and need to be simple enough for easy understanding by both domain experts and layman. Therefore, the analysis of the explanations is important since it moves us away from vague claims about interpretability and towards evaluating methods by a common set of terms [4]. There are three evaluation approaches for XAI being (i) functionally-grounded, where some formal definition of interpretability is used as a proxy for explanation quality analysis, (ii) application-grounded, where human experts evaluate the quality of the explanations in the context of the end-task, and (iii) human-grounded, where lay human-subject experiments are carried out to test the explanation quality regardless of its correctness [4, 27].

The majority of the papers that do perform evaluation adopt one of the last two evaluation approaches, making use of human subjects as their evaluators. However, as noted by [46], evaluating interpretable systems using only human evaluations can imply a strong bias towards simpler descriptions that might not completely represent the underlying reasoning of the method. In this project, we address this gap by adopting a comprehensive evaluation approach, where apart from the usual human subjective analysis, the interpretability efficiency is also analysed through functionally-grounded techniques so as to provide results in non-subjective and scientific metrics. Since this type of evaluation approach is rarely used throughout the XAI literature, it is difficult to compare to existing XAI techniques and interpretable systems in terms of such metrics [10]. Table 2 lists the hypotheses (A-F) that are tested during each of the three analysis approaches.

Table 2. The hypotheses tested by each XAI evaluation approach.

|   | Hypothesis Description | Functionally-grounded | Application-grounded | Human-grounded |
|---|---|---|---|---|
| A | The explanations provided are complete and correct | ✓ | ✓ | |
| B | The explanations provided are effective and useful | | ✓ | ✓ |
| C | The explanations provided are easily understood | | ✓ | ✓ |
| D | The explanations provided boost trust in ML models | | ✓ | ✓ |
| E | The explanations provided are sufficiently detailed | | ✓ | ✓ |
| F | Different explanations are required by different people | | ✓ | |

The below subsections include the evaluation results of each of the three types of analysis approaches on the explanations provided.

### 4.1. Functionally-Grounded Analysis

There are two types of functionally-grounded measures being complexity-based that analyses the complexity of the rule base, and semantics-based that analyses whether the semantics of the rules associate with the membership function. Motivated by Martens et al. [38] and the taxonomy proposed in [39], the below functionally-grounded measures are used in this study:

- **Number of unique rules**
- **Average number of rule conditions**
- **Consistency of rules** by checking if any contradicting rules exist and if rules are similar for instances of the same class
- **Completeness/Fidelity** by computing the percentage of instances where the model and the rules agree on the label

One might suggest that a lower number of rules are preferred since it makes it easier to follow through and keep track of things. However, more information in an explanation can also help users build a better mental model [40]. The consistency of the rules determines whether the method is trustworthy and reliable in its logic, whereas the completeness of the rules identifies how well they explain the decision function of the model, testing Hypothesis A. In the below subsections, analysis of each explanation type is carried out using the defined measures.

#### 4.1.1. Global Explanations

As discussed in Section 3.3.1, the global explanation is provided in the form of a decision tree, but for better comparison with the BRCG global explanation, the IF-THEN rules were also extracted from the tree. For BRCG, satisfying the DNF rule signifies that the loan application is likely to default, whilst for the implemented SHAP+GIRP method, satisfying either one of the IF-THEN rules results in a specific default rate.

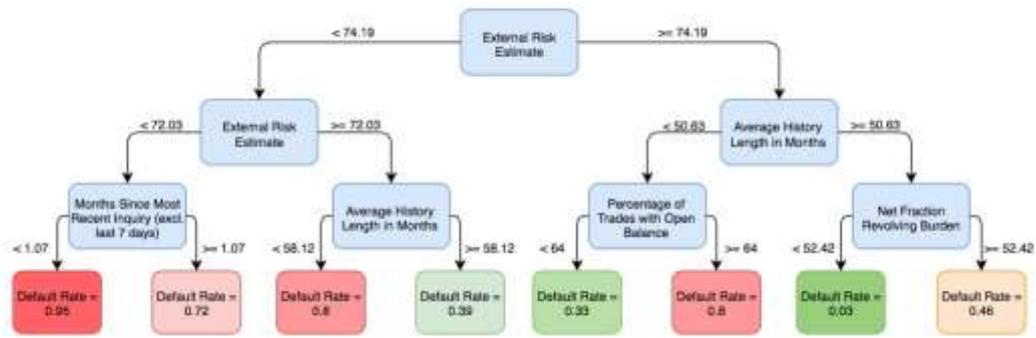

Figure 3. Global explanation for credit scoring model via SHAP + GIRP on HELOC Dataset.

Figure 3 illustrates the global explanation for the implemented XGBoost model on the HELOC Dataset, given in the form of a decision tree. As shown, when the *External Risk Estimate* feature has a smaller value, the loan application has a higher default risk. Moreover, a smaller *Percent Trades W Balance* leads to a lower default rate. These observations are inline with the monotonicity constraints of the HELOC Dataset. Table 3 depicts the evaluation results for both the BRCG and SHAP+GIRP global explanations on both datasets. For BRCG, each clause in the DNF rule (i.e. the rule of ANDs) is considered as a separate rule since either one of these clauses must be satisfied to satisfy the whole DNF rule. As shown, the BRCG model outputs fewer rules than the implemented SHAP+GIRP model. Impressively, BRCG manages to explain all the model predictions with just two rules of two conditions each for the HELOC Dataset and just one rule of two conditions for the LC Dataset. This could be possible due to BRCG's greedy approach in generating the rules. For the implemented SHAP+GIRP approach, the number and complexity of the rules could be easily adjusted by the maximum tree depth. Using a maximum tree depth of 2 resulted in 4 rules of 2 conditions each. As later shown during the application-grounded analysis (Section 4.2) having more rules and features could lead to more justifiable explanations that lead to more trust. In fact, most of the domain experts interviewed in this work suggested that having more features would have improved the explanations.

With regards to consistency, it is demonstrated in Table 3 that the global explanation provided by both models contains no conflicting and/or contradicting rules. This shows that the global explanation provided for XGBoost by the implemented SHAP+GIRP method is reliable and logical. When considering the completeness metric (i.e. the fidelity of the explanation to the predictions), the global explanation provided by the BRCG method is almost 100% complete for both datasets. BRCG achieves this high completeness rate because it is intrinsically explainable. On the other hand, the global explanation of the implemented XGBoost model achieves a completeness rate of around 90% on the HELOC Dataset and around 97% for the Lending Club Dataset. The loss in fidelity is most probably due to the extrinsic nature of the XAI method since the explanation was extracted from the XGBoost model through two levels of external processes, firstly SHAP to extract the feature contributions and then GIRP to form the global interpretation tree using these contributions. Therefore, it seems reasonable for such an explanation to have a lower completeness rate than an explanation extracted from a natively interpretable method. It is interesting to note that when increasing the maximum tree depth of the GIRP method to 100 on the HELOC Dataset, as done for other applications in [21], the completeness rate increased slightly by around 1%. However, the interpretability of the model was then greatly diminished since the decision tree became very hard to follow and the number of rules and conditions increased immensely. Therefore, completeness must be slightly sacrificed for considerably better interpretability. All in all, the completeness rate achieved by the SHAP+GIRP method is still quite high for both datasets, confirming Hypothesis A.

Table 3. Functionally-grounded metrics on global explanations.

|   | Metric | BRCG | SHAP + GIRP |
|---|---|---|---|
| HELOC | Number of unique rules | 2 | 8 |
| HELOC | Average number of rule conditions | 2 | 3 |
| HELOC | Consistency of rules | Yes | Yes |
| HELOC | Completeness rate | 99.96% | 89.95% |
| LC | Number of unique rules | 1 | 8 |
| LC | Average number of rule conditions | 2 | 3 |
| LC | Consistency of rules | Yes | Yes |
| LC | Completeness rate | 99.66% | 96.88% |

### 4.1.2. Local Feature-Based Explanations

As discussed in Section 3.3.2, the local feature-based explanations provided by Anchors are given in the form of DNF rules that contain conditions on the most important features. Since BRCG does not provide local feature-based explanations, these explanations cannot be compared to the benchmark.

|  | Example 1 | Example 2 |
|---|---|---|
| installment | 689.89 | 762.08 |
| int_rate | 14.65 | 8.9 |
| loan_amnt | 20000 | 24000 |
| credit_age | 5021 | 6180 |
| delinq_2yrs | 0 | 0 |
| inq_last_6mths | 2 | 1 |
| revol_util | 77.4 | 21.9 |
| open_acc | 14 | 12 |
| annual_install_to_inc | 0.15 | 0.13 |
| revol_bal_to_mnth_inc | 3.88 | 1.63 |
| term | 36 months | 36 months |
| sub_grade | F5 | A2 |
| home_ownership | MORTGAGE | OWN |

**Explanation for Example 1:**

Predicted as **Bad** because:
home_ownership = MORTGAGE
AND sub_grade = F5
AND annual_install_to_inc > 0.12
AND revol_bal_to_mnth_inc > 3.20

**Explanation for Example 2:**

Predicted as **Good** because:
home_ownership = OWN
AND annual_install_to_inc > 0.04
AND sub_grade = A2

Figure 4. Two examples of the local feature-based explanations via Anchors on LC Dataset

Figure 4 illustrates two examples of the provided local feature-based explanations from the Lending Club Dataset. As opposed to the global explanation, a local explanation is data-point specific. Therefore, the number of rules is equal to the size of the test set. As shown in Table 4 the average number of rule conditions is 4 for both datasets. This is quite a reasonable number of conditions since it provides sufficient details without complicating the rule too much. Moreover, the local feature-based explanations provided are consistent and 100% complete for each dataset since each rule is faithful to the data point and prediction it is explaining, confirming Hypothesis A.

Table 4. Functionally-grounded metrics on local feature-based explanations.

|  | Metric | Anchors |
|---|---|---|
| H | Average number of rule conditions | 4 |
| H | Consistency of rules | Yes |
| H | Completeness rate | 100% |
| LC | Average number of rule conditions | 4 |
| LC | Consistency of rules | Yes |
| LC | Completeness rate | 100% |

### 4.1.3. Local Instance-Based Explanations

As discussed in Section 3.3.3, the local instance-based explanations provided by ProtoDash are given in the form of prototypical instances from the data. This type of explanation assumes that loan applications that exhibit similar behaviours might end up in the same situation. It is important to note that BRCG does not provide local instance-based explanations.

Figure 5 illustrates an example local instance-based explanation provided for the LC Dataset. The *Loan application* column lists the feature values for the application at hand, whereas the two other columns list the feature values along with the target class and prototype weight of the two prototypical samples extracted as local instance-based explanations. Identical feature values are highlighted in green.

|  | Loan application | Prototype A | Prototype B |
|---|---|---|---|
| Target Class |  | Good | Good |
| Prototype Weight |  | 52% | 48% |
| installment | 445.1 | 186.61 | 587.71 |
| int_rate | 7.07 | 7.49 | 7.12 |
| loan_amnt | 14400 | 6000 | 19000 |
| credit_age | 4687 | 25933 | 6178 |
| delinq_2yrs | 0 | 0 | 0 |
| inq_last_6mths | 0 | 0 | 0 |
| revol_util | 33 | 88.2 | 44.6 |
| open_acc | 12 | 13 | 14 |
| annual_install_to_inc | 0.08 | 0.03 | 0.13 |
| revol_bal_to_mnth_inc | 3.66 | 14.88 | 5.16 |
| term | 36 months | 36 months | 36 months |
| sub_grade | C1 | D1 | A5 |
| home_ownership | MORTGAGE | MORTGAGE | MORTGAGE |

Figure 5. Sample local instance-based explanation via ProtoDash on LC Dataset.

Similar to the local feature-based explanations, such explanations are data-point specific, such that the prediction of each instance in the test set is explained through the use of two prototypical instances from the training set. It is noted that functionally-grounded analysis on local instance-based explanations in terms of number of rules, rule conditions and consistency is meaningless since these explanations are not given in the form of rules. As proven in this work and as stated in [37], local explanations are more faithful than global explanations.

### 4.2. Application-Grounded Analysis

As stated in [37], there is a lack of formalism on how this type of XAI evaluation, or any type of XAI evaluation for that matter, must be performed. Application-grounded analysis requires human domain experts to quantify the correctness and quality of the explanations provided by performing real tasks. In credit scoring, loan officers are considered experts the area since they have comprehensive knowledge of loan requirements and banking regulations.

In this project, interviews were carried out with seven different loan and/or risk officers employed at different banking and financial institutions around Malta (Bank of Valletta, HSBC, APS, BNF, Lombard). Each interview was around an hour long. The authors in [41] state that 5-10 respondents are needed to get reasonably stable psychometric estimates for evaluating the communality of answers. Application-grounded evaluation helps to identify the actual impact of the proposed model in a real-world application, since it directly tests the objective that the system was built for, giving strong indication on the actual success. To keep the duration of the

Computer Science & Information Technology (CS & IT)                                    197

interviews as short as possible, the evaluation was performed for just the HELOC Dataset. During the interviews, the domain experts were presented with a total of 3 tasks; one for each explanation type (global, local feature-based, and local instance-based). It is important to note that throughout the interview, it was observed that the interviewees' limited knowledge on the dataset contributed to an undesirable decrease in understandability. Therefore, it is worthy to mention that if the questions could make use of a dataset that the experts are used to and confident with, their understandability would have been certainly improved.

### 4.2.1. Global Explanations

Table 5 depicts the evaluation results achieved for each question from this section. As a general note, the *Result* column represents the literal result of the question, the *Percentage* column represents the result in the form of a global percentage, whilst the *Hypothesis* column lists the hypotheses confirmed by each question.

Table 5. Evaluation results acquired from interviews on the global explanation.

| Question/Task Description | Result | Percentage | Hypothesis |
| --- | --- | --- | --- |
| Forward prediction task | 7/7 experts | 100% | A & C |
| Accept/reject loan task | 7/7 experts | 100% | A & C |
| Preference of tree or rules? | 6/7 experts | 86% | - |
| How well explanation clarifies prediction? | 31/35 | 89% | B |
| Is explanation sufficiently detailed? | 6/7 experts | 86% | E |
| How much explanation increased trust in ML models? | 26/35 | 74% | D |

Firstly, the domain experts were presented with a forward prediction task where they were requested to interpret the model's prediction given the global explanation. 100% of the domain experts managed to correctly complete this task and reach the same conclusion as the model, confirming the understandability and correctness of the explanation. For the second task, the experts were requested to use the model's prediction and explanation to indicate whether they would accept or reject the loan application. Despite their limited dataset knowledge, all seven experts agreed with the model's prediction. Both these tasks confirm Hypotheses A and C.

Interestingly, six out of seven experts prefer the decision tree representation of the explanation. This implies that most humans find graphical representations more interpretable and easier to follow. Confirming this observation, the authors in [37] state that visualisations are the most human-centred technique for interpretability. This suggests that the global explanation provided as a decision tree might be preferred to BRCG's global explanation provided as a DNF rule, even though the BRCG explanation is simpler when comparing the rule-form of both methods.

Confirming Hypothesis B, the domain experts indicate that they believe that the explanation adequately clarifies the prediction, marking the Likert scale with an average score of 4.5. Moreover, six out of seven experts indicate that they are satisfied with the level of detail provided by the global explanation, confirming Hypothesis E. Finally, despite being quite a controversial question since people outside the AI community might find it hard to trust ML models, the domain experts indicate that the global explanation increased their trust by an average of 74%, confirming Hypothesis D.

### 4.2.2. Local Feature-Based Explanations

Table 6 depicts the results achieved for each question from this section. All the seven domain experts agreed with the model's prediction when asked to accept/reject a loan application, determining that the explanation provided is correct and easy to understand, thus confirming Hypothesis A and C. Regarding the explanation's ability to clarify the prediction, a total score of 29/35 (i.e. 83%) and an average score of 4.1 was achieved, which is slightly lower than the score achieved for this same question on the global explanation, but still confirms Hypothesis B. Six out of seven experts indicate that the local feature-based explanation provided is sufficiently detailed, confirming Hypothesis E. Finally, it is shown that on average, the local feature-based explanation increased the trust of the domain experts in ML models by 77%, which is slightly better than the score acquired for the global explanations. This confirms Hypothesis D. From the general impression given by the experts and as later described in Section 4.2.4, it was observed that most of the experts preferred the local feature-based explanation over the global explanation. An interesting point that was highlighted by one of the experts is that complete trust in the model is not necessarily required since the model should be there to help rather than make the decisions itself. Therefore, the ability to understand the reasoning behind the model's decision should provide enough trust to use the model.

Table 6. Evaluation results acquired from interviews on the local feature-based explanation.

| Question/Task Description | Result | Percentage | Hypothesis |
|---|---|---|---|
| Accept/reject loan task | 7/7 experts | 100% | A & C |
| How well explanation clarifies prediction? | 29/35 | 83% | B |
| Is explanation sufficiently detailed? | 6/7 experts | 86% | E |
| How much explanation increased trust in ML models? | 27/35 | 77% | D |

### 4.2.3. Local Instance-Based Explanations

Table 7 depicts the results achieved for the local instance-based explanations. Similar to the other two types of explanations, all seven domain experts agreed with the model's prediction when asked to accept/reject a loan application, confirming Hypotheses A and C. With regards to the explanation's ability to clarify the prediction, an average score of 3.3 is achieved, suggesting that this type of explanation was not as favoured and possibly implies that local instance-based explanations are not as effective and useful as the other two types. Most of the experts also specified that having three or four prototypes as part of the explanation, instead of two, would have been more useful. This could be easily resolved by adjusting the number of prototypes outputted from ProtoDash. Finally, it is shown that the local instance-based explanation increased the trust of the domain experts in ML models by an average of 74%. This is the same score achieved for the global explanation, which is slightly lower than that achieved for the local feature-based explanation. Having said this, the results confirm Hypothesis D. In general, whilst a few of the experts liked the idea behind the prototypical explanations, two experts expressed their disagreement with comparing loan applications to each other. They state that every case is different even if they show similar traits. Moreover, unpredictable changes can cause loan applications with very good traits to default and hence comparing with such application can cause unreliable results.

Table 7. Evaluation results acquired from interviews on the local instance-based explanation.

| Question/Task Description | Result | Percentage | Hypothesis |
|---|---|---|---|
| Accept/reject loan task | 7/7 experts | 100% | A & C |
| How well explanation clarifies prediction? | 23/35 | 66% | B |
| Is explanation sufficiently detailed? | 6/7 experts | 86% | E |
| Two prototypes are enough? | 2/7 experts | 30% | - |
| How much explanation increased trust in ML models? | 26/35 | 74% | D |

### 4.2.4. Final Thoughts

Finally, each domain expert was asked to choose their preferred type(s) of explanation(s). Some of the domain experts specified that all the explanation types are meaningful in different setups and recommended that all the three types should be available since a user might require a second form of explanation to confirm what is understood from the first explanation. The most preferred type of explanation is the local feature-based rule explanation, whilst the local instance-based explanation is the least preferred. However, the fact that each type of explanation was selected as a preferred explanation by one or more experts confirms that not everybody prefers the same thing. Therefore, this confirms Hypothesis F and thus affirms the efforts in this work with regards to providing three types of explanations for better subjective interpretability.

### 4.3. Human-Grounded Analysis

Human-grounded analysis includes conducting simpler human-subject experiments that still maintain the importance of the target application. Such evaluation can be carried out by lay humans, allowing for a bigger subject pool. This analysis focuses mainly on evaluating the quality and interpretability of the explanations rather than the correctness to ensure that the provided explanations are interpretable not just by domain experts but even lay humans.

In this project, human-grounded analysis is performed through questionnaires, which were sent over to a number of subjects of different age, gender, occupation, education level and marital status. Authors in [42] and [43] suggest that 10 to 30 participants are an adequate sample size. For the analysis, a Google Form was posted on a number of Facebook groups with members having different backgrounds, and 100 participants have completed the questionnaire. To keep the questionnaire as simple as possible, the evaluation was performed for just the HELOC Dataset. The participants are given a case scenario, where they are asked to imagine themselves as a loan applicant that has been denied a loan and has been provided with the model explanation for their denial. It is important to note that some of the features for the loan application were removed, whilst the rest were given in easy terms to keep the task simple and easy to complete. Since it is said that loan applicants prefer explanations that are related to their own case, only a feature-based explanation is used for this analysis as the participants are representing the loan applicants (which are lay human) in real life. The participants are asked to fill out a total of 5 questions using Likert scales, yes/no selection and textual answers.

Through the human-grounded analysis, some interesting observations were made. Firstly, 87% of the participants (i.e. participants that marked Question 1 with a score of 3, 4, or 5) were satisfied with the local feature-based explanation provided and, on average, the participants were 74% satisfied with the explanations. Moreover, 89% of the participants (i.e. participants that marked Question 2 with a score of 3, 4, or 5) found the explanation to be profitably understandable with the explanation achieving an average understandability of 78% amongst all the participants. The explanation provided also helped 17% of the participants to be have 100% more trust in ML models, whilst 38% of the participants were convinced to have more trust in ML models with 80% confidence. On average, with the help of the local feature-based explanation, the participants were 70% convinced to have more trust in AI models. This question is rather controversial since lay humans may have less knowledge on ML and AI and might therefore find it harder to trust such models. It is assumed that trustworthiness in such AI and ML models will increase with time as their use continues to expand and the models continue to improve in terms of interpretability [44, 45]. Furthermore, 72% of the participants found the explanation to be sufficiently detailed, whilst others suggested that more features, an overall risk rating or visualisation charts should be added. All in all, these results further confirm that the local feature-based explanations satisfy Hypotheses B-E.

## 5. CONCLUSIONS

In this work, a credit scoring model with state-of-the-art classification performance on the HELOC and Lending Club Datasets and comparable explainability to the benchmark BRCG model by Dash et al. [25] is proposed. The implemented credit scoring model incorporates the XGBoost algorithm, which demonstrates its capability of keeping a good balance between Type-I and Type-II errors. Furthermore, in aim of boosting the explainability of the black-box XGBoost model, a 360-degree explanation framework is developed by augmenting three separate post-hoc XAI techniques to provide three different types of explanations. A SHAP+GIRP method provides global explanations, Anchors provides local feature-based explanations and ProtoDash provides local instance-based explanations. Changing the classification function requires no changes in the interpretability component of the proposed model since the implemented XAI methodologies are model-agnostic and can be extracted from the current system pipeline and appended to a new classifier. It is shown, through the functionally-grounded analysis, that all the types of explanations provided are simple, consistent and complete. With regards to global explanations, it is shown that the provided explanation is comparatively as simple as the explanation produced by the benchmark BRCG model (in terms of number of rules and rule conditions). The application-grounded analysis deduced that six out of seven domain experts preferred the visual representation of the provided global explanation, which further suggests that the provided global explanation (in the form of a decision tree) might be preferred over the DNF rule of the BRCG model. The two other types of explanations are implemented over and above the global explanation and enable the implemented credit scoring model to be explained in alternative forms. In fact, the results of the application-grounded analysis present that the most preferred type of explanations are the local feature-based explanations, which are not provided by the benchmark BRCG model. It was also concluded that most of the financial experts interviewed found the explanations provided to be useful and have potential to be implemented in the system adopted by their bank. Through the rest of the evaluation, it was shown that the three types of explanations provided are complete and correct, effective and useful, easily understood, sufficiently detailed and trustworthy.

Future work can be focused on implementing a more user-specific solution with capabilities that allow the user to manage parameters such as the decision tree depth of the global explanations, the number of features in the local feature-based explanations, and the number of instances in the local instance-based explanations. Moreover, a possible improvement to the explanations, which was a popular suggestion amongst the domain experts interviewed, is to combine the global and local explanations in order to generate a decision tree that provides both global and local reasoning by highlighting the path and leaf node that is satisfied by the loan application in question. Future works suggested by lay humans through the human-grounded analysis include adding an overall risk rating or a percentage of eligibility to the explanation.

Computer Science & Information Technology (CS & IT) 201

**ACKNOWLEDGEMENTS**

The authors would like to gratefully thank Andrew Borg and BRSAnalytics for their financial support. Moreover, many thanks go to Susan Vella, Maria Grech, Rosalie Galea, Anthony Bezzina, Ian Cunningham, Jeremy Aguis and Mariella Vella for their help and time with the interviews, as well as the participants that took the time to fill out the questionnaire.

## AUTHORS

**Lara Marie Demajo** has degrees in M.Sc. in Artificial Intelligence and B.Sc. in Information Technology (Hons.) in AI, both from University of Malta. Her work has won various prizes including IEEE Best ICT Project and first place in FICTeX Final Year Project awarded by the Dean of Faculty of ICT. She has over 4 years experience in software development.

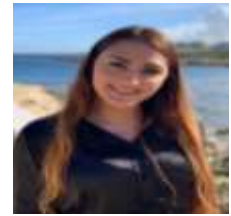

**Dr Vince Vella** brings over 25 years of senior technical leadership and management experience. Currently, he holds the position of CTO at Computime Software, BRSAnalytics and CTLabs. He holds a PhD from the Centre for Computational Finance and Economic Agents (CCFEA), University of Essex. Vince is also a lecturer within the Department of AI at University of Malta, mainly responsible for the MSc AI – Fintech stream. His main interests overlap Artificial Intelligence, Machine Learning and Computational Finance, particularly in the areas of AI Managed Funds, Algorithmic Trading, decentralized AI and AI for Anti Money Laundering.

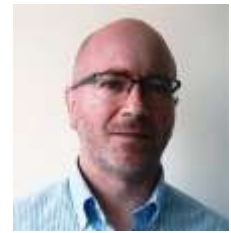

**Prof Alexiei Dingli** is a Professor of Artificial Intelligence (AI) at the Department of AI within the University of Malta. He has been conducting research and working in the field of AI for the past two decades. His work was rated World Class by international experts and won various prizes including; the Semantic Web Challenge, the first prize by the European Space Agency, the e-Excellence Gold Seal award, the First Prize in the Malta Innovation Awards, the World Intellectual Property Organization (WIPO) award for Creativity and the first prize of the Energy Globe award by the UN, amongst others. He has published several peer-reviewed papers and books in the field. He also formed part of the Malta.AI task-force aimed at making Malta one of the top-AI countries in the world where he chaired the working-group on AI in work & education. Prof Dingli also assists various local and international organizations during their transformation towards becoming AI companies.

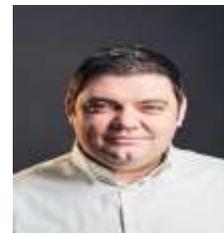